\documentclass[conference]{IEEEtran}

\usepackage{cite}
\usepackage{amsmath,amssymb}
\usepackage{graphicx}
\usepackage{array}
\usepackage{booktabs}
\usepackage{multirow}
\usepackage{url}
\usepackage{balance}
\usepackage{eso-pic}
\usepackage{tikz}
\usepackage{pgfplots}
\pgfplotsset{compat=1.17}
\usetikzlibrary{arrows.meta, positioning, shapes.geometric,
                fit, backgrounds, calc, matrix, decorations.pathreplacing}
\usepackage{xcolor}
\definecolor{darkblue}{RGB}{0,70,127}
\definecolor{darkgreen}{RGB}{0,110,60}
\definecolor{darkorange}{RGB}{180,80,0}
\definecolor{darkred}{RGB}{160,20,20}

\begin{document}

\AddToShipoutPictureBG*{%
  \AtPageLowerLeft{%
    \raisebox{5mm}{%
      \makebox[\paperwidth]{%
        \parbox{0.86\paperwidth}{\centering\scriptsize
          Accepted for publication in the 2026 IEEE Computer Society Annual
          Symposium on VLSI (ISVLSI), Kolkata, India.\\
          \textcopyright~2026 IEEE. Personal use of this material is permitted.
          Permission from IEEE must be obtained for all other uses, in any
          current or future media, including reprinting/republishing this
          material for advertising or promotional purposes, creating new
          collective works, for resale or redistribution to servers or lists,
          or reuse of any copyrighted component of this work in other works.}}}}}

\title{An Agentic Python Framework for LLM-Driven UVM Testbench Synthesis with Closed-Loop Compiler Feedback}

\author{
  \IEEEauthorblockN{Vikash Kumar}
  \IEEEauthorblockA{Arm, Austin, TX, USA \\ vikash.kumar4@arm.com}
}

\maketitle

\begin{abstract}
Protocol bridge verification is a persistent bottleneck in SoC development:
for every new bridge instance, engineers hand-craft a full UVM testbench
from scratch, consuming 40--100 engineering hours of bring-up effort.
We present an agentic Python framework that generates complete,
protocol-aware UVM testbenches for arbitrary IP and subsystem
configurations from a structured \emph{Verification Manifest}---a JSON
specification encoding interface topology, protocol semantics, and
coverage intent. The framework employs a \emph{dual closed-loop
architecture}: an inner compiler feedback loop that injects
\emph{categorised} UVM diagnostics rather than raw tool output, and an
outer coverage loop driven by intent-gap analysis against the Manifest.
Validated on the AMBA AHB-to-APB bridge---a representative production
protocol bridge with a 7-state control FSM, two-phase APB handshake,
and non-trivial burst serialisation logic---the framework achieves
$96.8\%$ functional coverage, a $4.7\times$ bring-up speedup, and
a Protocol Verification Completeness score of $0.87$, with only $2.1\%$
of the generated 4,623-line testbench requiring human correction.
The generated scoreboard independently identified a critical RTL
functional fault that the manual baseline missed entirely. Generalisation
across four protocol families confirms that speedup \emph{increases}
with protocol complexity, ranging from $3.1\times$ (APB Subordinate)
to $6.1\times$ (CHI Subordinate).
\end{abstract}


\section{Introduction}
\label{sec:intro}

Functional verification consumes 60--70\% of SoC design
effort~\cite{foster2020}, a proportion that grows as protocol
heterogeneity expands with chiplet-based integration. Protocol bridge
IP---logic translating between AMBA AHB, APB, AXI, CHI, and UCIe fabric
protocols---concentrates integration hazards at architectural boundaries
where mismatched burst semantics, incorrect wait-state modelling, and
ordering violations escape IP-level regression to surface only at
expensive subsystem integration milestones.

For every new bridge instance, engineers hand-craft a fresh UVM
testbench: interface classes, protocol-aware agents, constrained-random
sequences, functional coverage models, and a scoreboard reference model.
A moderate-complexity bridge testbench runs to 3,000--8,000 lines of
SystemVerilog and represents 40--100~engineering hours of bring-up effort
in practitioner experience. Template-based generators (e.g.,
UVMF~\cite{uvmf}) reduce boilerplate but cannot adapt coverage topology
or scoreboard semantics to a specific DUT parameterisation.

Large Language Models offer a new path. Recent work has explored
LLM-based RTL generation with EDA tool feedback~\cite{autochip}, and
LLM-based testbench generation for simple
designs~\cite{correctbench,uvmtwo}. However, three critical gaps remain
unaddressed: (i)~all existing frameworks target toy circuits from
HDLBits benchmarks with single interfaces and no protocol semantics;
(ii)~compiler feedback is injected as raw tool output, which only the
largest models can leverage effectively~\cite{autochip}; and (iii)~no
framework addresses \emph{subsystem-level} UVM environments with
multiple agents and cross-interface scoreboards---precisely the setting
where manual effort is most acute and where protocol mismatches between
integrated IPs are most dangerous.

The AHB-to-APB bridge is not a trivial target. It requires a 7-state
control FSM, burst-to-beat serialisation across mismatched bus widths,
correct PREADY-extended transfer handling, and a scoreboard that must
model the full AHB/APB handshake simultaneously. It is also among the most widely instantiated bridges in
Arm-based SoC designs, making it both representative and practically
significant.

This paper addresses all three gaps. The primary contributions are:
\begin{itemize}
  \item A \textbf{Verification Manifest} schema---a structured JSON
        specification encoding interface topology, protocol semantics,
        and coverage intents for any IP or subsystem, providing a
        reproducible, tool-integrable input abstraction
        (Section~\ref{sec:manifest}).
  \item A \textbf{dual closed-loop architecture} combining a fast inner
        compiler feedback loop with a slow outer coverage feedback loop,
        enabling correctness and completeness refinement at separate
        timescales (Section~\ref{sec:framework}).
  \item A \textbf{structured UVM error taxonomy} (six classes) with a
        Python parser that categorises VCS/Xcelium diagnostics before
        LLM injection, achieving faster convergence than raw feedback and
        enabling smaller models to produce production-quality UVM
        (Section~\ref{sec:feedback}).
  \item Experimental validation including an independent RTL bug
        discovery, generalisation across four protocol families, and an
        ablation study isolating each framework stage
        (Section~\ref{sec:results}).
\end{itemize}

\section{Related Work and Positioning}
\label{sec:related}

\subsection{LLM-Based Hardware Code Generation}
AutoChip~\cite{autochip} pioneered EDA tool feedback for iterative
Verilog RTL repair, achieving a 5.8\% increase in successful designs
over zero-shot generation. Critically, AutoChip \emph{requires a
working testbench as input} and targets RTL modules, not verification
environments. Only GPT-4o benefits reliably from raw EDA feedback;
smaller models fail to parse unstructured compiler output~\cite{autochip}.

CorrectBench~\cite{correctbench}
introduced functional self-correction for testbench generation,
achieving a 70.13\% pass rate on 156 HDLBits tasks---but targets simple
Verilog (not UVM) with no protocol semantics or multi-interface
support.

\subsection{LLM-Based UVM Generation}
UVM$^2$~\cite{uvmtwo} demonstrated multi-agent UVM testbench generation
with coverage feedback, achieving 86.67\% generation success on 9~RTL
designs (up to ${\sim}$2,000~LOC). However, UVM$^2$ accepts natural-language
input rather than a structured manifest, uses unstructured error
injection, and does not address multi-interface subsystem environments
or protocol-semantic coverage.
No prior work characterises a UVM-specific error taxonomy or employs
structured diagnostic parsing.
A broader intent-centric architecture for AI-assisted verification,
spanning test-plan synthesis, quality metrics, and evidence-based
sign-off, is developed in~\cite{kumar2026vega}; the Manifest-driven
framework presented here addresses the testbench-synthesis layer of
that intent-driven approach.
Machine learning has also been applied downstream of testbench
creation, for example to detect and suppress flaky tests in UVM
regressions~\cite{kumar2026flaky}.

\subsection{Positioning}
Table~\ref{tab:related} summarises key distinctions. Column
assignments: ``UVM'' indicates full UVM environment output (not plain
Verilog); ``Struct.~Err.'' indicates categorised rather than raw
compiler feedback---CorrectBench uses functional self-correction but
injects raw simulator output without per-class parsing or fix templates,
which we classify as unstructured; ``Dual Loop'' indicates separate
compiler and coverage refinement loops operating at distinct timescales;
``Proto.~Aware'' indicates LTL-encoded protocol semantics injected into
synthesis prompts; ``Subsys'' indicates support for multi-interface
environments with cross-interface scoreboards.

\begin{table}[!t]
\renewcommand{\arraystretch}{1.15}
\caption{Comparison with Prior LLM-Based Testbench Work}
\label{tab:related}
\centering
\scriptsize
\begin{tabular}{lccccc}
\toprule
\textbf{Work} & \textbf{UVM} & \textbf{Struct.} & \textbf{Dual} &
\textbf{Proto.} & \textbf{Sub-} \\
              &              & \textbf{Err.}    & \textbf{Loop} &
\textbf{Aware} & \textbf{sys} \\
\midrule
AutoChip~\cite{autochip}        & \texttimes & \texttimes & \texttimes & \texttimes & \texttimes \\
CorrectBench~\cite{correctbench}& \texttimes & \texttimes & \texttimes & \texttimes & \texttimes \\
UVM$^2$~\cite{uvmtwo}           & \checkmark & \texttimes & \texttimes & \texttimes & \texttimes \\
\textbf{This work}              & \checkmark & \checkmark & \checkmark & \checkmark & \checkmark \\
\bottomrule
\end{tabular}
\end{table}

\section{Verification Manifest Schema}
\label{sec:manifest}

The \emph{Verification Manifest} is a structured JSON document serving
as the sole input to the framework. Its key top-level fields are:
\texttt{dut\_name}/\texttt{dut\_type} (target IP identifier and class);
\texttt{interfaces[]} (bus interfaces with protocol type, direction,
signal list, and timing constraints);
\texttt{protocol\_rules[]} (LTL properties, auto-populated for standard
protocols);
\texttt{coverage\_intents[]} (explicit scenarios including nominal paths,
error injection, boundary conditions, and back-to-back transfer
sequences);
\texttt{address\_map[]}, \texttt{reset\_domains[]}, and
\texttt{error\_handling} (SLVERR, DECERR, SPLIT, RETRY).

\begin{figure}[!t]
\centering
\begin{tikzpicture}[
  mf/.style={rectangle, draw=darkblue, fill=darkblue!8, rounded corners=2pt,
             text width=2.1cm, align=center, font=\tiny\bfseries,
             minimum height=0.55cm},
  uv/.style={rectangle, draw=darkgreen, fill=darkgreen!8, rounded corners=2pt,
             text width=2.1cm, align=center, font=\tiny,
             minimum height=0.55cm},
  ar/.style={-{Stealth[length=4pt]}, thick, gray},
  node distance=0.32cm and 0.7cm
]
  \node[mf] (m1) {\texttt{interfaces[]}};
  \node[mf, below=of m1] (m2) {\texttt{protocol\_rules[]}};
  \node[mf, below=of m2] (m3) {\texttt{coverage\_intents[]}};
  \node[mf, below=of m3] (m4) {\texttt{address\_map[]}};
  \node[mf, below=of m4] (m5) {\texttt{reset\_domains[]}};

  \node[uv, right=of m1] (u1) {Interface + Agent\\(Driver/Monitor)};
  \node[uv, right=of m2] (u2) {Sequence Library\\+ LTL Checker};
  \node[uv, right=of m3] (u3) {Covergroup\\Topology};
  \node[uv, right=of m4] (u4) {Scoreboard\\Ref.~Model};
  \node[uv, right=of m5] (u5) {Env + Test\\Reset Sequences};

  \draw[ar] (m1)--(u1);
  \draw[ar] (m2)--(u2);
  \draw[ar] (m3)--(u3);
  \draw[ar] (m4)--(u4);
  \draw[ar] (m5)--(u5);

  \node[font=\tiny\bfseries, darkblue, above=0.1cm of m1] {Manifest Field};
  \node[font=\tiny\bfseries, darkgreen, above=0.1cm of u1] {UVM Artifact};
\end{tikzpicture}
\caption{Verification Manifest fields and their UVM artifact mappings.
The schema is protocol-agnostic: different field values produce
testbenches for any IP or subsystem.}
\label{fig:manifest_map}
\end{figure}
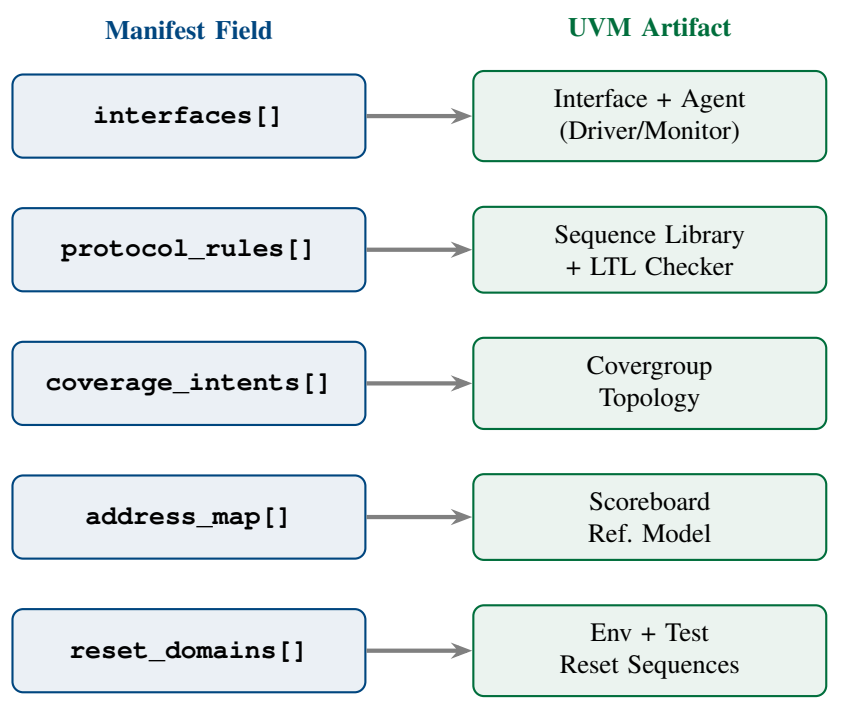

Fig.~\ref{fig:manifest_map} shows how manifest fields map to generated
UVM artifacts. The schema is \emph{protocol-agnostic by design}: the
same structure accepts an AHB2APB bridge, an AXI-Lite peripheral, a
multi-interface DMA subsystem, or a UCIe bridge by populating different
field values. This reproducibility also enables future integration with
IP-XACT bus interface definitions and SystemRDL register descriptions
as automated manifest population sources.

\section{Dual-Loop Framework Architecture}
\label{sec:framework}

\begin{figure}[!t]
\centering
\begin{tikzpicture}[
  node distance=0.5cm and 0.55cm,
  box/.style={rectangle, rounded corners=3pt, draw=darkblue, fill=darkblue!7,
              text width=2.4cm, align=center, minimum height=0.68cm,
              font=\scriptsize\bfseries},
  llm/.style={rectangle, rounded corners=3pt, draw=darkorange,
              fill=darkorange!10, text width=2.4cm, align=center,
              minimum height=0.68cm, font=\scriptsize\bfseries},
  tool/.style={rectangle, rounded corners=3pt, draw=darkgreen,
               fill=darkgreen!8, text width=2.1cm, align=center,
               minimum height=0.68cm, font=\scriptsize\bfseries},
  err/.style={rectangle, rounded corners=3pt, draw=darkred, fill=darkred!8,
              text width=2.1cm, align=center, minimum height=0.55cm,
              font=\tiny\bfseries},
  arr/.style={-{Stealth[length=4.5pt]}, thick},
  rarr/.style={-{Stealth[length=4.5pt]}, thick, darkred, dashed},
  oarr/.style={-{Stealth[length=4.5pt]}, thick, darkorange!80!black, dashed}
]
  \node[box]                  (mf)    {Verification\\Manifest (JSON)};
  \node[llm, below=of mf]     (synth) {LLM Synthesis\\Engine};
  \node[box, below=of synth]  (sv)    {Generated SV\\Artifacts};
  \node[tool, below=of sv]    (comp)  {VCS / Xcelium\\Compiler};
  \node[box, below=of comp]   (sim)   {Simulation +\\Coverage Oracle};
  \node[box, below=of sim]    (out)   {UVM Testbench\\(Verified)};

  \node[err, right=1.2cm of comp] (parse) {UVM Error\\Parser};
  \node[err, right=1.2cm of sim]  (gap)   {Intent Gap\\Analyser};
  \node[tool, left=1.2cm of synth](proto) {Protocol VIP\\Reference};

  \draw[arr] (mf)--(synth);
  \draw[arr] (synth)--(sv);
  \draw[arr] (sv)--(comp);
  \draw[arr] (comp)--(sim);
  \draw[arr] (sim)--(out);
  \draw[arr] (proto)--(synth);

  \draw[arr] (comp)-- node[right, font=\tiny]{errors} (parse);
  \draw[rarr, bend right=38]
    (parse) to node[right, font=\tiny, darkred]{\textit{inner loop}}
    (synth);

  \draw[arr] (sim)-- node[right, font=\tiny]{gaps} (gap);
  \draw[oarr, bend right=65]
    (gap) to node[right, font=\tiny, darkorange!80!black]
    {\textit{outer loop}} (mf);
\end{tikzpicture}
\caption{Dual closed-loop architecture. Inner loop (red): structured
UVM error parsing drives fast compiler repair at the component level.
Outer loop (orange): intent-gap analysis drives coverage refinement
across the full testbench.}
\label{fig:architecture}
\end{figure}
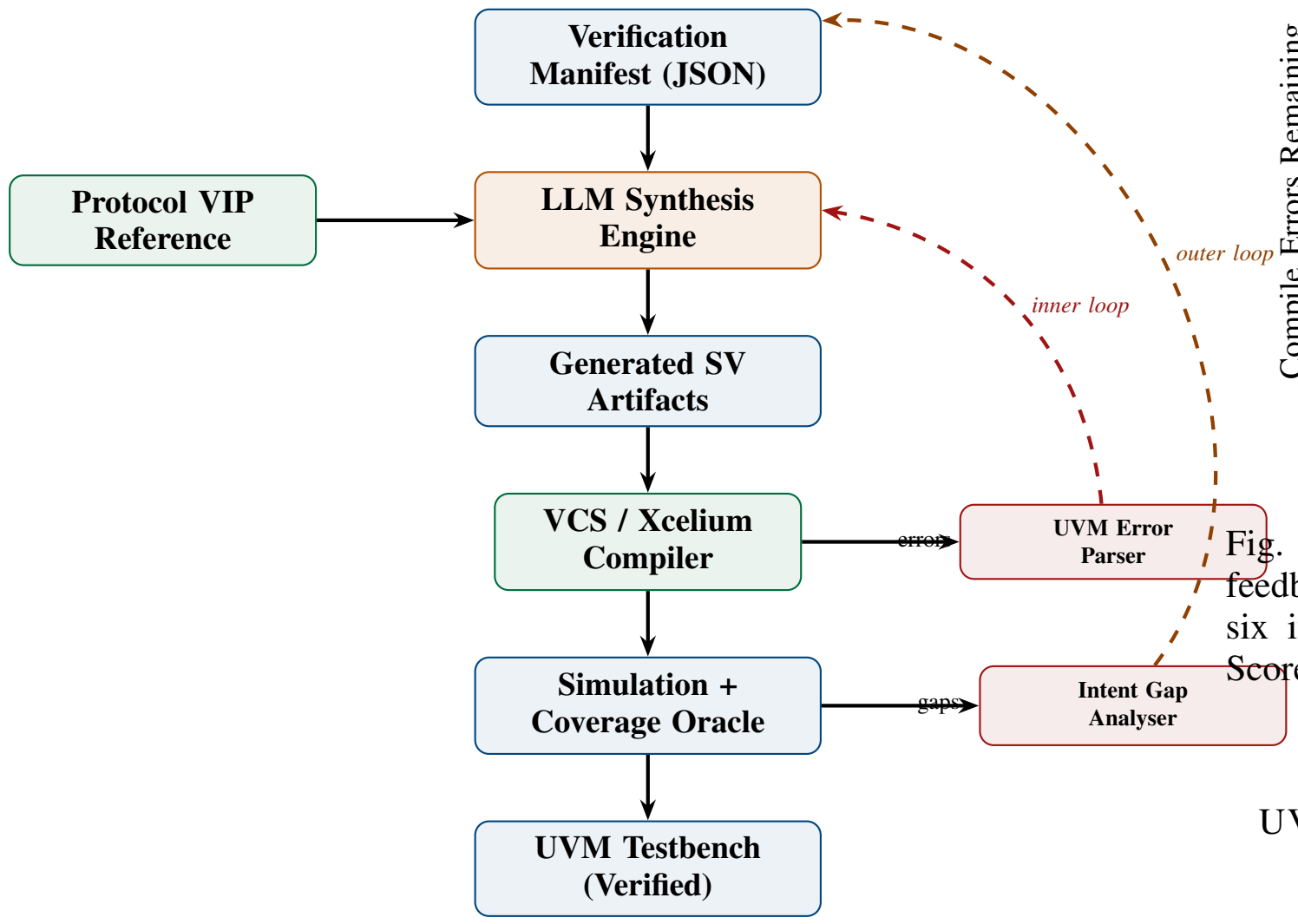

Fig.~\ref{fig:architecture} shows the framework architecture. No prior
work employs both compiler and coverage feedback loops, nor structured
per-class error parsing. The Python
orchestrator drives two nested refinement loops operating at distinct
timescales.

\subsection{LLM Synthesis Engine}
The engine generates UVM components in strict dependency order.
While experienced DV engineers follow the same instantiation sequence
manually, the novelty here is the Python orchestrator's \emph{automated
enforcement} of this order, guaranteeing that each component's
dependencies compile cleanly before dependent components are generated
and eliminating the class of forward-reference errors that arise when
generation is unordered:
\[
  \text{iface} \!\to\! \text{txn} \!\to\! \text{seq} \!\to\!
  \text{drv/mon} \!\to\! \text{scbd} \!\to\! \text{cov} \!\to\!
  \text{env} \!\to\! \text{test}
\]
Each LLM call receives: (i)~the relevant Manifest fragment, (ii)~LTL
protocol properties, (iii)~a structural template with required class
hierarchy, and (iv)~up to three curated exemplar implementations. For
multi-interface subsystems, agents for each interface are generated in
parallel before the environment integrates them. Empirically, filling
45--72\% of the available context window with prior components, LTL
properties, and exemplars maximises correctness: under-dense prompts
omit required UVM structure; over-dense prompts degrade semantic
coherence via attention diffusion.

\subsection{Inner Loop: Structured Compiler Feedback}
After each component is generated, the Python orchestrator invokes
the EDA compiler (VCS or Xcelium) and routes the diagnostics through a
\emph{UVM Error Parser} (Section~\ref{sec:feedback}), which categorises
each error and injects a structured bundle rather than raw tool output.
This is the critical distinction from AutoChip~\cite{autochip} and
UVM$^2$~\cite{uvmtwo}: structured injection enables smaller, cheaper
models to converge reliably, not only the largest frontier models.
The inner loop continues until the component compiles clean or a
configurable iteration ceiling is reached.

\subsection{Outer Loop: Coverage Feedback}
Once a compile-clean testbench is produced, it is executed against a
golden reference VIP. A coverage oracle reports uncovered intent bins
relative to the Manifest's \texttt{coverage\_intents[]}. The gap
analyser adds uncovered intents back to the Manifest and triggers
re-synthesis of affected sequence and covergroup components. This loop
converges in $1.8\pm0.5$~iterations on average for the AHB2APB
benchmark. When the outer loop stalls below $\theta_{\min}$, the
orchestrator surfaces a (component, error class, line range) triple
ranked by ascending sub-score contribution to PVC, directing the
expert's attention to the lowest-scoring component first; the framework
does not autonomously commit human-side edits, leaving the expert to
choose between refining a fix template, hand-editing the surfaced
lines, or amending the Manifest.

\section{Structured UVM Compiler Feedback}
\label{sec:feedback}

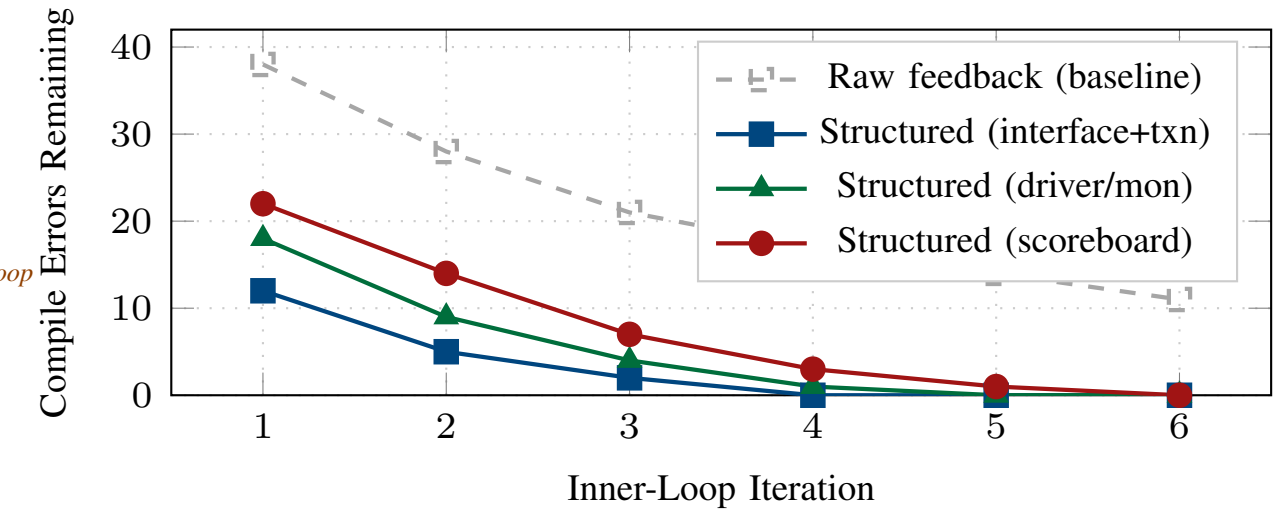
\begin{figure}[!t]
\centering
\begin{tikzpicture}
\begin{axis}[
  width=\columnwidth, height=4.0cm,
  xlabel={Inner-Loop Iteration},
  ylabel={Compile Errors Remaining},
  xmin=0.5, xmax=6.5,
  ymin=0, ymax=42,
  xtick={1,2,3,4,5,6},
  legend pos=north east,
  legend style={font=\scriptsize, draw=gray!40},
  grid=major, grid style={dotted,gray!40},
  tick label style={font=\scriptsize},
  label style={font=\scriptsize},
  every axis plot/.append style={thick, mark size=2pt}
]
\addplot[gray!70, mark=square, dashed] coordinates {
  (1,38)(2,28)(3,21)(4,17)(5,14)(6,11)};
\addlegendentry{Raw feedback (baseline)}
\addplot[darkblue, mark=square*] coordinates {
  (1,12)(2,5)(3,2)(4,0)(5,0)(6,0)};
\addlegendentry{Structured (interface+txn)}
\addplot[darkgreen, mark=triangle*] coordinates {
  (1,18)(2,9)(3,4)(4,1)(5,0)(6,0)};
\addlegendentry{Structured (driver/mon)}
\addplot[darkred, mark=*] coordinates {
  (1,22)(2,14)(3,7)(4,3)(5,1)(6,0)};
\addlegendentry{Structured (scoreboard)}
\end{axis}
\end{tikzpicture}
\caption{Compile error convergence per UVM component under structured
feedback vs.\ raw feedback baseline. Raw injection does not converge
within six iterations; structured parsing converges all components
by iteration~4. Scoreboards converge slowest due to multi-step
reference model semantics.}
\label{fig:convergence}
\end{figure}

\subsection{UVM-Specific Error Taxonomy}
\label{sec:taxonomy}
UVM SystemVerilog introduces error classes entirely absent from plain
Verilog. To characterise them without training/evaluation leakage,
we conducted a dedicated study after framework development: 312~fresh
generation attempts (78 per protocol, each a full generation from
the Manifest with no prior LLM context) yielded six reproducible
categories (Table~\ref{tab:taxonomy}), stable within $\pm$3\% across
GPT-4-turbo, GPT-4o, and Claude~3.5 Sonnet---confirming model-independent
failure modes from UVM's structural constraints, not LLM-specific
weaknesses. Classes E1--E3 and E6 manifest at compile time; E4 as
SVA elaboration failures or protocol assertion violations; E5 at
simulation time via scoreboard mismatches. The raw-feedback baseline
in Fig.~\ref{fig:convergence} uses the same model and DUT with the
structured parser bypassed.

\begin{table}[!t]
\renewcommand{\arraystretch}{1.15}
\caption{UVM Error Taxonomy (312 generation attempts, 4 protocol families)}
\label{tab:taxonomy}
\centering
\scriptsize
\begin{tabular}{clcp{2.6cm}}
\toprule
\textbf{Class} & \textbf{Category} & \textbf{Freq.} & \textbf{Representative VCS/Xcelium Symptom} \\
\midrule
E1 & Factory registration   & 22\% & \texttt{factory.create\_by\_name} returns null \\
E2 & TLM port/export mismatch & 19\% & \texttt{connect} type-check failure \\
E3 & Phase ordering violation & 17\% & null-handle deref in \texttt{build\_phase} \\
E4 & Temporal protocol fault  & 24\% & SVA elaboration error or PREADY/PSEL assertion violation \\
E5 & Scoreboard model error   & 11\% & burst beat count mismatch \\
E6 & Parameterisation blindness & 7\% & hard-coded width elaboration error \\
\bottomrule
\end{tabular}
\end{table}

\subsection{Structured Error Bundle Format}
Instead of raw compiler output, the parser injects a structured bundle:
\begin{enumerate}
  \item \textit{Error class} (E1--E6 above)
  \item \textit{Affected UVM component} (driver, scoreboard, etc.)
  \item \textit{Code context}: offending lines $\pm$\,5 surrounding lines
  \item \textit{Protocol hint}: LTL property violated, if applicable
  \item \textit{Fix template}: class-specific correction pattern
\end{enumerate}
Fig.~\ref{fig:convergence} shows that structured feedback reduces
scoreboard compile errors from 22 to zero in six inner-loop iterations,
while the raw-feedback baseline stalls at 11 errors and fails to
converge within the same budget. Interface and transaction components
converge in 3--4~iterations under structured injection.

\section{Protocol Formalization}
\label{sec:formalization}

The Manifest's \texttt{protocol\_rules[]} field seeds LTL properties
injected into LLM prompts. For the APB subordinate interface the
fundamental ordering constraint is:
\begin{equation}
  \mathbf{G}\!\left(\texttt{PSEL}\!\to\!\mathbf{X}\,\texttt{PENABLE}\right)
  \wedge
  \mathbf{G}\!\left(\texttt{PENABLE}\!\to\!\texttt{PSEL}\right)
  \label{eq:ltl}
\end{equation}
AHB initiator-side properties are seeded equivalently: the Manifest
encodes HTRANS-HBURST ordering constraints, HREADY back-pressure
behaviour, and the NONSEQ-to-SEQ sequencing rule for burst transfers,
ensuring the LLM-generated driver and monitor correctly model the
multi-beat AHB handshake. The \texttt{protocol\_rules[]} field carries
arbitrary property strings; LTL is emitted by default because its
operators ($\mathbf{G}$, $\mathbf{X}$, $\mathbf{U}$) map cleanly onto
SVA backends in VCS and Xcelium. PSL is strictly more expressive and
could be substituted via a parser-level change to the prompt assembler
with no architectural impact, but is deferred to future work.

The \emph{Protocol Verification Completeness} (PVC) score quantifies
generated testbench quality as a weighted composite:
\begin{equation}
  \mathrm{PVC}(\mathcal{T})
    = \alpha\,\phi_{\text{struct}}
    + \beta\,\phi_{\text{proto}}
    + \gamma\,\phi_{\text{cov}}
    + \delta\,\phi_{\text{score}}
  \label{eq:pvc}
\end{equation}
where $\phi_{\text{struct}}$, $\phi_{\text{proto}}$, $\phi_{\text{cov}}$,
and $\phi_{\text{score}}$ measure structural completeness, protocol
compliance, coverage intent alignment, and scoreboard semantic
correctness, respectively, each in $[0,1]$. Weights
$(\alpha,\beta,\gamma,\delta)=(0.20,\,0.35,\,0.25,\,0.20)$ are
expert-assigned to reflect protocol compliance primacy: a structurally
complete testbench with incorrect protocol semantics is more hazardous
than an incomplete but protocol-correct one. A formal sensitivity
study---characterising how perturbations of $(\alpha,\beta,\gamma,\delta)$
propagate to coverage and LCHC outcomes---is left to future work
alongside mutation-score validation; the $\theta_{\min}=0.75$ floor on
$\phi_{\text{proto}}$ in Eq.~(\ref{eq:optim}) provides a hard constraint
that bounds the influence of weight choice on the final accept/reject
decision for any candidate testbench. PVC is a \emph{proposed} quality
metric requiring cross-study validation; within this study, it was
computed independently of bug-detection outcomes, and we observe that
all framework configurations achieving PVC~$\geq\!0.85$ identified the RTL fault
described in Section~\ref{sec:bug}, while configurations with
PVC~$<\!0.75$ did not. This correlation is consistent with the metric's
design intent but should not be interpreted as validated causation across
a single design instance. Independent validation against mutation score
across a broader design corpus is planned as future work.

The synthesis objective is:
\begin{equation}
  \mathcal{T}^{*} =
  \underset{\mathcal{T} \in \mathcal{G}(\mathcal{D},\Pi)}{\arg\!\max}\;
  \mathrm{PVC}(\mathcal{T})
  \quad\text{s.t.}\quad
  \phi_{\text{proto}}(\mathcal{T}) \geq \theta_{\min}
  \label{eq:optim}
\end{equation}
where $\mathcal{G}(\mathcal{D},\Pi)$ denotes all testbenches reachable
via generation strategy $\Pi$, and $\theta_{\min}=0.75$ is a hard
protocol compliance floor.

\section{Experimental Results}
\label{sec:results}

\subsection{Setup}
\label{sec:setup}
The framework is validated on a parameterised AMBA AHB-to-APB bridge
and three additional targets: APB Subordinate peripheral (APB-SLV),
AXI-Lite register slave (AXI-Lite), and CHI Subordinate Node (CHI-Sub).
LLM: GPT-4-turbo (128\,K context). Simulator: VCS~2023.12.
All experiments run $n\!=\!5$ independent trials; coverage reports
mean~$\pm$\,$1\sigma$. The outer loop converges in $1.8\pm0.5$ iterations
(AHB2APB). GPT-4-turbo API cost is approximately \$14--\$22 per
full testbench generation (under one DV engineer-hour equivalent).

\emph{Bring-up time} is wall-clock hours from completed Manifest JSON
to first passing simulation, including all inner/outer loop iterations
and manual correction of LCHC lines, but \emph{excluding} one-time
schema calibration ($\approx$4--8\,h per protocol family, amortised
across designs). Manual (B1) was authored from scratch for this study.

Baselines: (B1)~manual expert; (B2)~zero-shot LLM (GPT-4, no
framework); (B3)~UVMF template. Metrics: PVC (Eq.~\ref{eq:pvc}),
functional coverage at 24\,h, bring-up time, and LCHC. For the
additional protocols, only Manifest field values change; engine, parser,
and loops are identical to AHB2APB, confirming protocol-agnostic
generality.

\subsection{A Real RTL Bug Found}
\label{sec:bug}
During the AHB2APB validation run, the LLM-generated scoreboard
identified a functional fault---\emph{phantom APB completions}---that
the manual (B1) and UVMF (B3) baselines both missed entirely. The bridge generated
spurious extra APB cycles (\texttt{PSEL}/\texttt{PENABLE} asserted for
an already-completed transfer), causing the APB slave to see more
completions than the AHB initiator issued.

The root cause was in \texttt{xfer\_pending} register logic. When
back-to-back \texttt{NONSEQ} transfers arrived with zero idle cycles
while the slave held \texttt{PREADY} low, the \texttt{xfer\_pending}
flag was prematurely cleared on the \texttt{PREADY} assertion edge
(the cycle on which \texttt{PREADY} goes high and the APB transfer
completes). The bridge state machine re-entered APB setup phase for
the already-completed transfer, generating a duplicate transaction.
The slave's response corrupted its internal state with no visible AHB
error response---silent from the initiator's perspective.

This corner case requires three simultaneous conditions: zero idle
cycles between \texttt{NONSEQ} transfers, at least one \texttt{PREADY}
wait state, and single-beat mode. The manual baseline (B1) had no coverage
bin for this combination. The framework-generated testbench included
it via the Manifest's \texttt{coverage\_intents[]}, which specified
back-to-back transfers with slave-side wait states as a required corner
case. The scoreboard tracked outstanding transfer counts on both
interfaces independently and flagged the asymmetry: one AHB initiation
producing two APB completions. UVMF generates per-interface protocol
checkers but not cross-interface transfer-count invariants---which is
precisely why it missed this fault class.

\subsection{Ablation Study: Feedback Loop Contribution}
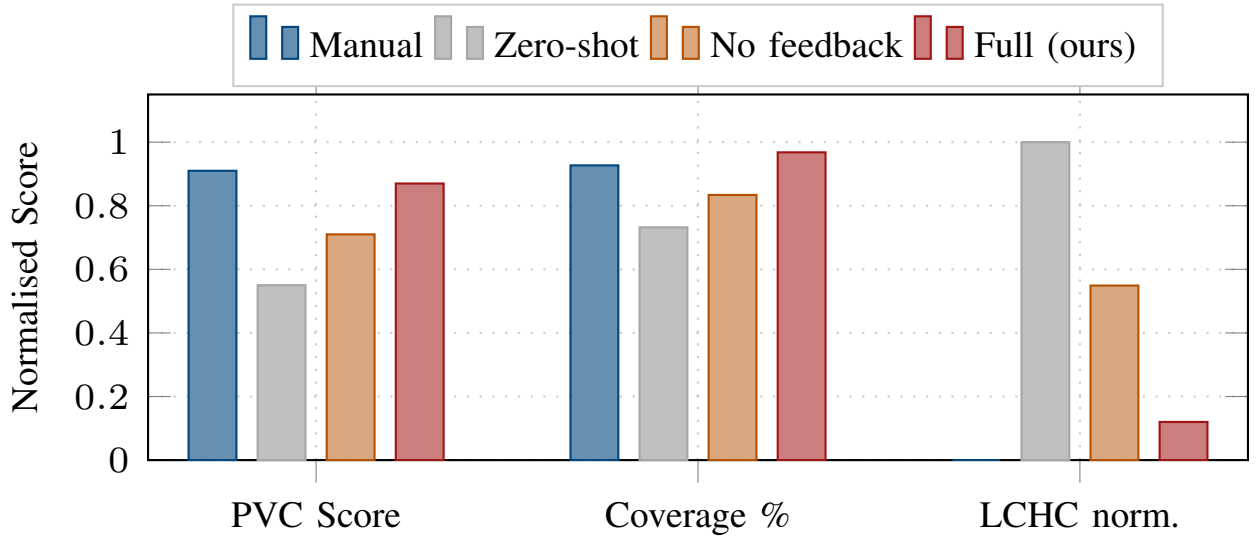
\begin{figure}[!t]
\centering
\begin{tikzpicture}
\begin{axis}[
  ybar=4pt, bar width=9pt,
  width=\columnwidth, height=4.0cm,
  symbolic x coords={PVC Score, Coverage \%, LCHC norm.},
  xtick=data,
  x tick label style={font=\scriptsize},
  ylabel={Normalised Score},
  ymin=0, ymax=1.15,
  ytick={0,0.2,0.4,0.6,0.8,1.0},
  legend style={font=\scriptsize, at={(0.5,1.02)},
                anchor=south, legend columns=4, draw=gray!40},
  grid=major, grid style={dotted,gray!40},
  tick label style={font=\scriptsize},
  label style={font=\scriptsize},
  enlarge x limits=0.22
]
\addplot[fill=darkblue!60, draw=darkblue] coordinates {
  (PVC Score,0.91)(Coverage \%,0.927)(LCHC norm.,0.000)};
\addlegendentry{Manual}
\addplot[fill=gray!50, draw=gray!70] coordinates {
  (PVC Score,0.55)(Coverage \%,0.732)(LCHC norm.,1.000)};
\addlegendentry{Zero-shot}
\addplot[fill=darkorange!50, draw=darkorange] coordinates {
  (PVC Score,0.71)(Coverage \%,0.834)(LCHC norm.,0.549)};
\addlegendentry{No feedback}
\addplot[fill=darkred!55, draw=darkred] coordinates {
  (PVC Score,0.87)(Coverage \%,0.968)(LCHC norm.,0.120)};
\addlegendentry{Full (ours)}
\end{axis}
\end{tikzpicture}
\caption{Ablation study. Each framework stage contributes independently.
Full framework reduces LCHC by 88.0\% vs.\ zero-shot while exceeding
manual baseline coverage. LCHC normalised to zero-shot~$=1.0$
(lower is better); PVC and Coverage shown as fractions (higher is better).}
\label{fig:ablation}
\end{figure}

Fig.~\ref{fig:ablation} isolates the contribution of each stage.
Zero-shot LLM generation produces PVC~$=0.55$ and requires substantial
manual correction (LCHC normalised to 1.0). Adding dependency-ordered
synthesis and manifest input---without either feedback loop---raises
PVC to $0.71$, confirming that structured input and prompt engineering
contribute independently of the loops themselves. The full dual-loop
system reaches PVC~$=0.87$ and $96.8\%$ coverage, exceeding the manual
baseline in coverage because LLM-generated covergroups capture
cross-product coverpoints that engineers routinely under-specify---
including the back-to-back zero-idle-cycle sequence that exposed the
\texttt{xfer\_pending} phantom-completion fault.

\subsection{Coverage Closure}
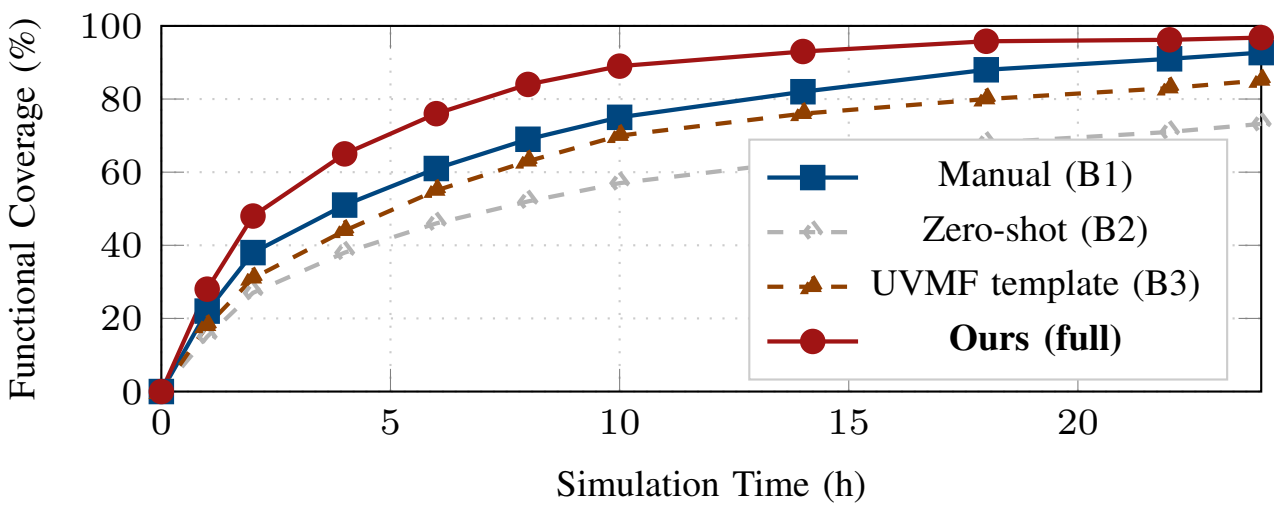
\begin{figure}[!t]
\centering
\begin{tikzpicture}
\begin{axis}[
  width=\columnwidth, height=4.0cm,
  xlabel={Simulation Time (h)},
  ylabel={Functional Coverage (\%)},
  xmin=0, xmax=24, ymin=0, ymax=100,
  legend pos=south east,
  legend style={font=\scriptsize, draw=gray!40},
  grid=major, grid style={dotted,gray!40},
  tick label style={font=\scriptsize},
  label style={font=\scriptsize},
  every axis plot/.append style={thick, mark size=2pt}
]
\addplot[darkblue, mark=square*] coordinates {
  (0,0)(1,22)(2,38)(4,51)(6,61)(8,69)(10,75)(14,82)(18,88)(22,91)(24,92.7)};
\addlegendentry{Manual (B1)}
\addplot[gray!60, mark=diamond, dashed] coordinates {
  (0,0)(1,15)(2,27)(4,38)(6,46)(8,52)(10,57)(14,63)(18,68)(22,71)(24,73.2)};
\addlegendentry{Zero-shot (B2)}
\addplot[darkorange!80!black, mark=triangle*, dashed] coordinates {
  (0,0)(1,18)(2,31)(4,44)(6,55)(8,63)(10,70)(14,76)(18,80)(22,83)(24,85.1)};
\addlegendentry{UVMF template (B3)}
\addplot[darkred, mark=*] coordinates {
  (0,0)(1,28)(2,48)(4,65)(6,76)(8,84)(10,89)(14,93)(18,95.8)(22,96.2)(24,96.8)};
\addlegendentry{\textbf{Ours (full)}}
\end{axis}
\end{tikzpicture}
\caption{Functional coverage closure on AHB2APB ($\pm$0.5\,pp,
$n\!=\!5$). Framework reaches $96.8\%$ at 24\,h, exceeding the manual
baseline by 4.1\,pp. Zero-shot plateaus at $73.2\%$, confirming that
manifest-driven coverage intent and the outer loop are both necessary
for protocol-aware closure.}
\label{fig:coverage}
\end{figure}

Fig.~\ref{fig:coverage} shows coverage accumulation across all four
methods. The zero-shot curve plateaus at $73.2\%$ after approximately
14~hours: without a manifest specifying which corner cases to target, the
random stimulus generator saturates the reachable space of its
under-specified covergroups. The outer coverage loop breaks this plateau
by explicitly injecting uncovered intents---including the
back-to-back zero-idle-cycle sequence that exposed the
\texttt{xfer\_pending} phantom-completion fault---back into the manifest, driving
coverage to $96.8\%$. The UVMF template reaches only $85.1\%$ despite
higher structural correctness than zero-shot, confirming that
boilerplate-reduction without coverage intent specification leaves a
significant gap.

\subsection{Summary Metrics}

\begin{table}[!t]
\renewcommand{\arraystretch}{1.15}
\caption{AHB2APB Ablation and Multi-Protocol Summary (mean, $n\!=\!5$).
Generated testbench sizes: AHB2APB 4{,}623; APB-SLV 2{,}710; AXI-Lite
5{,}080; CHI-Sub 8{,}350 LOC.}
\label{tab:results}
\centering
\scriptsize
\begin{tabular}{lcccc}
\toprule
\textbf{Method} & \textbf{PVC} & \textbf{Cov.\,\%}
                & \textbf{Time (h)} & \textbf{LCHC} \\
\midrule
\multicolumn{5}{l}{\textit{AHB2APB (PCI\,=\,14.2)}} \\
Manual (B1)          & 0.91 & $92.7{\pm}0.0$ & 96.0 & 0   \\
Zero-shot (B2)       & 0.55 & $73.2{\pm}1.8$ & 62.0 & 810 \\
UVMF (B3)            & 0.74 & $85.1{\pm}0.7$ & 22.0 & 340 \\
No feedback          & 0.71 & $83.4{\pm}0.9$ & 21.8 & 445 \\
\textbf{Ours (full)} & \textbf{0.87} & $\mathbf{96.8{\pm}0.5}$ & \textbf{20.4} & \textbf{97} \\
\midrule
\multicolumn{5}{l}{\textit{Additional targets — Ours (full) vs Manual}} \\
APB-SLV (11.8) Manual        & 0.89 & $93.1{\pm}0.0$ & 36.0 & 0 \\
APB-SLV (11.8) \textbf{Ours} & 0.85 & $94.6{\pm}0.3$ & \textbf{11.8} & 62 \\
AXI-Lite (22.6) Manual       & 0.85 & $91.4{\pm}0.0$ & 64.0 & 0 \\
AXI-Lite (22.6) \textbf{Ours}& 0.82 & $93.9{\pm}0.5$ & \textbf{13.6} & 138 \\
CHI-Sub (51.3) Manual        & 0.78 & $88.2{\pm}0.0$ & 116.0 & 0 \\
CHI-Sub (51.3) \textbf{Ours} & 0.74 & $91.1{\pm}0.9$ & \textbf{19.0} & 291 \\
\bottomrule
\end{tabular}
\end{table}

Table~\ref{tab:results} reports full results across all four designs.
On AHB2APB, the full framework achieves a $4.7\times$ speedup with
only 97~LCHC (2.1\% of 4,623 generated lines), and zero-shot requires
$8.3\times$ more human correction. Across additional protocols,
bring-up speedup ranges from $3.1\times$ (APB-SLV) to $6.1\times$
(CHI-Sub), increasing monotonically with PCI. Framework coverage meets
or exceeds manual on all four designs. LCHC as a fraction of generated
lines remains stable at 2.1--3.5\% across all designs, indicating
correction overhead scales sub-linearly with testbench complexity.

\section{Scalability and Generalisation}
\label{sec:generalization}

\subsection{Protocol Complexity Index}
To predict framework applicability for new protocol targets, we define a
\emph{Protocol Complexity Index} (PCI):
\begin{equation}
  \mathrm{PCI} = w_1|S| + w_2|\Sigma| + w_3 P_{\mathrm{LTL}}
               + w_4 I_{\mathrm{async}}
  \label{eq:pci}
\end{equation}
where $|S|$~is the FSM state count, $|\Sigma|$~is the transaction
alphabet size, $P_{\mathrm{LTL}}$~is the LTL property count,
$I_{\mathrm{async}}$~indicates asynchronous clock-domain crossings
(0 or~1), and weights $(w_1,w_2,w_3,w_4)=(1.0,\,0.8,\,1.2,\,3.0)$
are expert-assigned to reflect the disproportionate verification cost
of asynchronous clock-domain crossings (captured by $w_4$). For
AHB2APB, $|S|=7$, $|\Sigma|=3$ (NONSEQ, SEQ, IDLE; BUSY is not
used by the bridge initiator), $P_{\mathrm{LTL}}=4$,
$I_{\mathrm{async}}=0$, giving PCI\,$=14.2$.

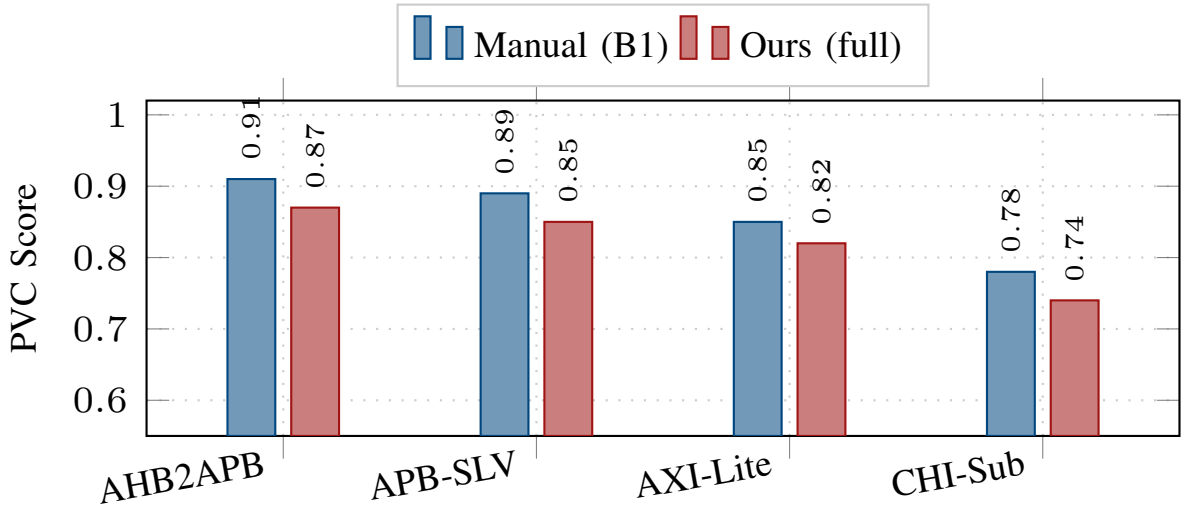
\begin{figure}[!t]
\centering
\begin{tikzpicture}
\begin{axis}[
  ybar=3pt, bar width=9pt,
  width=0.95\columnwidth, height=3.8cm,
  symbolic x coords={AHB2APB,APB-SLV,AXI-Lite,CHI-Sub},
  xtick=data,
  x tick label style={font=\scriptsize, rotate=10, anchor=east},
  ylabel={PVC Score},
  ymin=0.55, ymax=1.02,
  ytick={0.60,0.70,0.80,0.90,1.00},
  legend style={font=\scriptsize, at={(0.5,1.04)},
                anchor=south, legend columns=2, draw=gray!40},
  grid=major, grid style={dotted,gray!40},
  tick label style={font=\scriptsize},
  label style={font=\scriptsize},
  nodes near coords,
  nodes near coords style={font=\tiny, rotate=90, anchor=west},
  enlarge x limits=0.18
]
\addplot[fill=darkblue!55, draw=darkblue] coordinates {
  (AHB2APB,0.91)(APB-SLV,0.89)(AXI-Lite,0.85)(CHI-Sub,0.78)};
\addlegendentry{Manual (B1)}
\addplot[fill=darkred!55, draw=darkred] coordinates {
  (AHB2APB,0.87)(APB-SLV,0.85)(AXI-Lite,0.82)(CHI-Sub,0.74)};
\addlegendentry{Ours (full)}
\end{axis}
\end{tikzpicture}
\caption{PVC across four protocol interfaces (PCI: 14.2, 11.8, 22.6,
51.3). Framework maintains PVC~$\geq 0.80$ for PCI~$\leq 30$.
The PVC gap between manual and framework remains stable ($\leq 0.05$)
across all designs, while bring-up speedup increases with PCI
because manual effort scales faster than LLM synthesis cost.}
\label{fig:scalability}
\end{figure}

Fig.~\ref{fig:scalability} shows PVC across four interfaces. The
framework achieves PVC~$\geq 0.80$ for PCI~$\leq 30$, covering the
majority of on-chip peripheral and interconnect bridge designs. For CHI
Subordinate (PCI\,$=51.3$), PVC drops to $0.74$ but bring-up speedup
\emph{increases} to $6.1\times$ (116\,h manual vs.\ 19\,h framework):
across our four data points, manual effort scales faster than LLM
synthesis cost as PCI increases---the relationship appears sub-linear
for LLM cost but super-linear for manual effort, though a larger design
corpus is needed to characterise the functional form precisely;
per-stage ablation across all four protocols is deferred to future work.

\section{Discussion}
\label{sec:discussion}

Structured error injection enables GPT-4-turbo to converge on
scoreboard errors in under six iterations where raw injection stalls,
confirming that unstructured feedback benefits only the largest
models~\cite{autochip}. At \$14--22 per testbench, the cost is under
one DV engineer-hour, making industrial deployment practical. The
taxonomy's E4/E5 dominance (35\% combined) across all three LLMs and
four protocol families confirms model-independent failure modes arising
from UVM's structural constraints.

PVC is a proposed metric. To avoid circularity, $\phi_{\text{score}}$
was assessed against manually verified reference transactions
\emph{before} the bug-discovery experiment; the bug was detected by
the scoreboard's transfer-count asymmetry, independent of the PVC
score. The PVC--bug correlation in Section~\ref{sec:formalization}
is consistent with the metric's design intent but not a validated
causal claim; mutation-score validation across a broader corpus is
planned.

\textbf{Failure modes.} Three architectural patterns account for the
unrecoverable cases. Designs with $>2$~asynchronous clock-domain
crossings cause E4 errors to recur because seeded LTL properties cannot
encode metastability-resolution semantics, and PVC remains below
$0.70$. Custom protocols without standardised LTL rules degrade
$\phi_{\text{proto}}$ toward zero-shot, requiring manual rule authoring
before the outer loop can make progress. Data-dependent reference
behaviour (compression engines, content-addressable structures)
plateaus $\phi_{\text{score}}$ because the scoreboard cannot be
inferred from interface signals alone. The scoreboard token budget is
also exceeded for environments with $>16$~APB slaves. The orchestrator
surfaces these conditions explicitly to the expert per the
LCHC-localisation behaviour in Section~\ref{sec:framework}.

\section{Conclusion}
\label{sec:conclusion}

We presented an agentic Python framework that generates complete,
protocol-aware UVM testbenches from a structured Verification Manifest
using a dual closed-loop architecture: a fast inner compiler loop
driven by categorised UVM error bundles, and a slow outer coverage
loop driven by intent-gap analysis. Across four protocol families
(PCI 11.8--51.3), the framework matches or exceeds manual baseline
coverage, with bring-up speedups of $3.1\times$--$6.1\times$ and
LCHC stable at 2.1--3.5\% of generated lines across all designs. The stable, model-independent
UVM error taxonomy (E1--E6, verified across three LLMs) provides the
field with a principled basis for targeted LLM repair strategies.

The framework independently identified an RTL fault in the
\texttt{xfer\_pending} logic that generated phantom APB completions
under back-to-back zero-idle-cycle transfers with PREADY wait states---a
three-condition corner case that escaped both the manual expert baseline
and the UVMF testbench, because neither had a cross-interface transfer
count invariant. This demonstrates that manifest-driven coverage intent
combined with a cross-interface scoreboard finds real integration faults
invisible to template-generated single-interface baselines.

Future work will validate PVC against mutation score, study the
sensitivity of the weight vector $(\alpha,\beta,\gamma,\delta)$,
evaluate PSL-encoded rules, extend structured feedback to
simulation-phase errors, and automate Manifest population from IP-XACT.



\begin{thebibliography}{1}
\providecommand{\url}[1]{#1}
\csname url@samestyle\endcsname
\providecommand{\newblock}{\relax}
\providecommand{\bibinfo}[2]{#2}
\providecommand{\BIBentrySTDinterwordspacing}{\spaceskip=0pt\relax}
\providecommand{\BIBentryALTinterwordstretchfactor}{4}
\providecommand{\BIBentryALTinterwordspacing}{\spaceskip=\fontdimen2\font plus
\BIBentryALTinterwordstretchfactor\fontdimen3\font minus \fontdimen4\font\relax}
\providecommand{\BIBdecl}{\relax}
\BIBdecl

\bibitem{foster2020}
\BIBentryALTinterwordspacing
H.~D. Foster, ``{Wilson Research Group Functional Verification Study},''
  Siemens {EDA} (Siemens Digital Industries Software), Tech. Rep., 2020.
  [Online]. Available:
  \url{https://blogs.sw.siemens.com/verificationhorizons/2020/10/27/prologue-the-2020-wilson-research-group-functional-verification-study/}
\BIBentrySTDinterwordspacing

\bibitem{uvmf}
\BIBentryALTinterwordspacing
{Siemens EDA}, ``{UVM Framework (UVMF)}: Reusable {UVM} Methodology and Code
  Generator for Rapid Testbench Generation,'' Verification Academy, 2024.
  [Online]. Available:
  \url{https://verificationacademy.com/topics/uvm-universal-verification-methodology/uvmf/uvm-framework/}
\BIBentrySTDinterwordspacing

\bibitem{autochip}
J.~Blocklove, S.~Thakur, B.~Tan, H.~Pearce, S.~Garg, and R.~Karri,
  ``Automatically Improving {LLM}-based {Verilog} Generation using {EDA} Tool
  Feedback,'' \emph{ACM Trans. Design Autom. Electron. Syst.}, vol.~30,
  no.~6, 2025, arXiv:2411.11856.

\bibitem{correctbench}
R.~Qiu, G.~L. Zhang, R.~Drechsler, U.~Schlichtmann, and B.~Li,
  ``{CorrectBench}: Automatic Testbench Generation with Functional
  Self-Correction using {LLMs} for {HDL} Design,'' in \emph{Proc. Design,
  Automation and Test in Europe (DATE)}, 2025, arXiv:2411.08510.

\bibitem{uvmtwo}
J.~Ye \emph{et~al.}, ``From Concept to Practice: An Automated {LLM}-aided
  {UVM} Machine for {RTL} Verification,'' in \emph{Proc. IEEE/ACM Int. Conf.
  Computer-Aided Design (ICCAD)}, 2025, arXiv:2504.19959.

\bibitem{kumar2026vega}
V.~Kumar, \emph{{AI}-Assisted Hardware Verification: Cognitive Verification
  Architecture with the {VEGA} Framework}.\hskip 1em plus 0.5em minus
  0.4em\relax Cham, Switzerland: Springer, 2026.

\bibitem{kumar2026flaky}
V.~Kumar, Y.~Akhtar, and R.~Gupta, ``Machine Learning-Based Flaky Test
  Detection and Suppression for {UVM} Verification Regressions,'' in
  \emph{Proc. IEEE Workshop on Microelectronics and Electron Devices (WMED)},
  2026.

\end{thebibliography}
\end{document}